\documentclass[aps,pra,reprint,showpacs,twocolumn,groupedaddress]{revtex4-1}

\usepackage{graphicx}
\usepackage{dcolumn}
\usepackage{bm}
\usepackage{braket}
\usepackage{leftidx}
\usepackage{subcaption}
\usepackage{cancel}
\usepackage{amsmath}

\usepackage{color}
\usepackage[mathcal]{eucal}
\usepackage{epstopdf}

\begin{document}


\title{Conservation relation of nonclassicality and entanglement \\ for Gaussian states in a beam splitter}

\author{Wenchao Ge$^{1}$}\thanks{wenchaoge.tamu@gmail.com}
\author{Mehmet Emre Tasgin$^{1,2,3}$}\thanks{metasgin@hacettepe.edu.tr }
\author{M. Suhail Zubairy$^{1}$}\thanks{zubairy@physics.tamu.edu}

\address{$^{1}$Institute for Quantum Science and Engineering (IQSE) and Department of
Physics and Astronomy, Texas A\&M University, College Station, Texas 77843, USA\\
$^{2}$Institute of Nuclear Sciences, Hacettepe University, 06800, Ankara, Turkey \\
$^{3}$to whom correspondence shall be addressed}




\date{\today}

\begin{abstract}
We study the relation between single-mode nonclassicality and two-mode entanglement in a beam-splitter. We show that not all of the nonclassicality (entanglement potential) is transformed into two-mode entanglement for an incident single-mode light. Some of the entanglement potential remains as single-mode nonclassicality in the two entangled output modes. Two-mode entanglement generated in the process can be equivalently quantified as the increase in the minimum uncertainty widths (or decrease in the squeezing) of the output states compared to the input states. We use the nonclassical depth and logarithmic negativity as single-mode nonclassicality and entanglement measures, respectively. We realize that a conservation relation between the two quantities can be adopted for Gaussian states, if one works in terms of uncertainty width. This conservation relation is extended to many sets of beam-splitters.
\end{abstract}

\pacs{03.67.Bg, 03.67.Mn, 42.50.Dv, 42.50.Ex}

\maketitle
\section{Introduction}
Quantum entanglement is an essential source for quantum information processing \cite{Nielsen:10}. Entanglement between two-mode Gaussian states \cite{Wang:07, Braunstein:05} is of considerable interest because of its availability and controllability in experiment, and its applications such as quantum teleportation \cite{Braunstein:98} and dense coding \cite{Braunstein:00}.

Entanglement of two-mode Gaussian states can be generated in an experiment via a nonlinear optical device, such as a parametric down converter \cite{Kimble:87}. On the other hand, a beam splitter (BS) as a linear passive device has been studied extensively to generate quantum entanglement \cite{Tan:91, Sanders:92,Marian:01, Kim:02, Wolf:03, Li:06, Tahira:09,killoran2015converting,miranowicz2015statistical,brunelli2015single}. In particular, Kim \textit{et al.} \cite{Kim:02} studied the properties of different input states, such as squeezed states, in order to have the output fields to be entangled. They conjectured that nonclassicality of input fields is a necessary condition for entangling the output via a BS, which was proved by Wang \cite{Wang:02}. Wolf \textit{et al.} \cite{Wolf:03} proved a necessary and sufficient condition for entangling bipartite Gaussian states with passive optical devices and found a close relation between input squeezing and output entanglement of the Gaussian state. Furthermore, Tahira \textit{et al.} \cite{Tahira:09} investigated the generation of Gaussian entanglement from a single-mode squeezed state mixed with a thermal state at a BS, where detailed experimental conditions are analyzed.

Recently, a new measure for nonclassicality, entanglement potential, is introduced by Asb\'{o}th {\it et al.} \cite{Asboth:05} based on its inherent relation with the two-mode entanglement. Entanglement potential is the maximum amount of two-mode entanglement extractable from a single mode nonclassical state using linear optical devices. More recently, Vogel and Sperling \cite{Vogel:14} arrived at a more intimate connection between nonclassicality and two-mode entanglement. The rank of the two-mode entanglement that a nonclassical state can generate is equal to the number of terms needed in the coherent state expansion of this nonclassical state. It is also pointed out in \cite{Tasgin:15} that such a connection can exist in many-particle entanglement. With these results, an interesting question arises: is there a way to quantify single-mode nonclassicality and two-mode entanglement so that the summation of the two quantities is conserved via linear passive devices, such as a BS?

In this paper, we address this question for arbitrary two-mode Gaussian states. First, we show that not all of the nonclassicality, present in an input single-mode state, is converted to two-mode entanglement even for the optimum BS mixing angle (see Fig.~\ref{fig:figENvstau}). We calculate the remaining single-mode nonclassicalities by partially tracing the two output modes. This wipes out the two-mode entanglement and enables the calculation of single-mode nonclassical depths.

Second, we realize an interesting relation between the generated two-mode entanglement and the change in the logarithm of the initial and final nonclassical depths (or uncertainty widths). The change in the logarithm of the initial and final uncertainity widths [see Eq.~(\ref{eq:en_def})] displays the same behavior with the logarithmic negativity measure \cite{Vidal:02,adesso2004extremal,plenio2005logarithmic} for the generated two-mode entanglement (Fig.~\ref{fig:vacuum}). We use this observation for treating nonclassicality and entanglement with an equal footing, relying on the soul of the entanglement potential \citep{Asboth:05,Vogel:14}. Hence, a conservation relation for the sum of single-mode nonclassicality and generated two-mode entanglement can be deduced.

This paper is organized as follows. In Sec. II, we start with the basic theory of input and output states at a BS. In Sec.~IIB, we introduce the definition for the entanglement depth. In Sec. III, we show that nonclassicality in an input single-mode state cannot be extracted completely using only a single beam-splitter. When the two-mode entanglement is wiped out via partial trace operation, nonclassical depths of the output modes do not vanish. In Sec. IV, we outline the relation between the nonclassical depth and the uncertainty width of a single-mode state. We define the generated two-mode entanglement as the difference between input and output nonclasscialities in a BS. We show that this definition is qualitatively equivalent to the logarithmic negativity measure of entanglement. It enables us to convert the logarithmic negativity for the two-mode entanglement to single-mode nonclassicality and vice versa. A conservation relation of nonclassicality and entanglement is derived at a BS. We illustrate this relation with examples and extend the relation to many sets of BSs.  A summary of this paper is given in Sec. V. In the appendices, detailed derivations are provided.

\begin{figure}[t]
\centering
\includegraphics[width=0.425\textwidth]{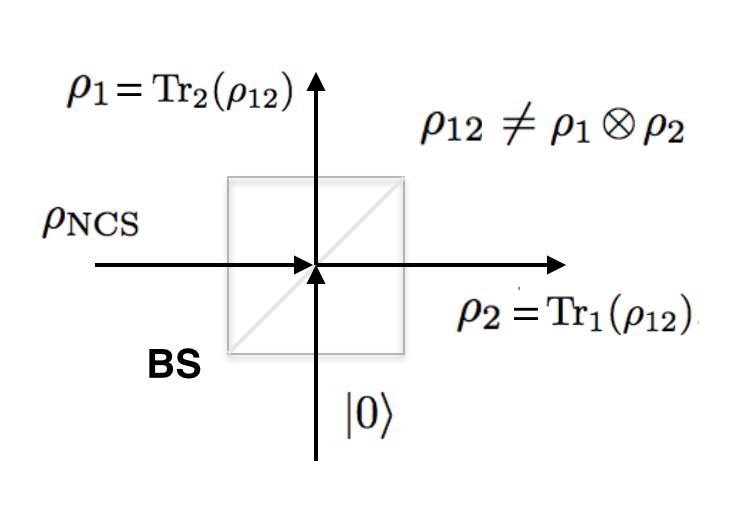}
\caption{\label{fig:scheme} (Color online). A nonclassical Gaussian state $\rho_{\text{NCS}}$ is mixed with a vacuum state at a BS, generating an output state $\rho_{12}$. Each output mode of the bipartite state after the BS is given by $\rho_1=\text{Tr}_2(\rho_{12})$ and $\rho_2=\text{Tr}_1(\rho_{12})$, respectively.}
\label{fig:scheme}
\end{figure}

\section{Input and output Gaussian states of a BS}
\subsection{Input-output relation}
We consider a lossless BS with two single mode Gaussian fields as input. The complex amplitudes $\beta_1,\beta_2$ of the output fields are related to those $\alpha_1, \alpha_2$ of the input fields as \cite{Campos:89}
\begin{eqnarray}
\left( \begin{array}{cc}
\beta_1\\
\beta_2
\end{array} \right)=M\left( \begin{array}{cc}
\alpha_1\\
\alpha_2
\end{array} \right),
\label{eq:BS}
\end{eqnarray}
where 
\begin{eqnarray}
M = \left( \begin{array}{cc}
\cos\theta & \sin\theta e^{i\varphi} \\
-\sin\theta e^{-i\varphi} & \cos\theta
\end{array} \right),
\label{eq:M}
\end{eqnarray}
is the beam splitter transformation matrix with the transmittance $\cos^2\theta$ and the phase difference $\varphi$ between the reflected and the transmitted fields.

For a single-mode Gaussian state, the characteristic function of the state is given by
\begin{eqnarray}
\chi(\alpha_i,\alpha_i^{\ast})=\exp\left(-\frac{1}{2}\boldsymbol{x}_i^{\dagger}V_i \boldsymbol{x}_i\right),
\end{eqnarray}
where $\boldsymbol{x}_i^{\dagger}=(\alpha_i^{\ast},\alpha_i)$, and $V_i$ is the covariance matrix of the single-mode state ($i=1,2$).
For two separable single-mode Gaussian states, the characteristic function of the states is given by
\begin{eqnarray}
\chi_{\text{in}}(\alpha_1,\alpha_1^{\ast},\alpha_2,\alpha_2^{\ast})&=&\chi(\alpha_1,\alpha_1^{\ast})\chi(\alpha_2,\alpha_2^{\ast})\nonumber\\
&=&\exp\left(-\frac{1}{2}\boldsymbol{y}^{\dagger}V_{\text{in}} \boldsymbol{y}\right),
\label{eq:chara_f}
\end{eqnarray}
where $\boldsymbol{y}^{\dagger}=(\alpha_1^{\ast},\alpha_1,\alpha_2^{\ast},\alpha_2)$ and $V_{\text{in}} = \left( \begin{array}{cc}
V_1 & 0 \\
0 & V_2
\end{array} \right)$ is the input covariance matrix. For a general input two-mode Gaussian $V_1=\left( \begin{array}{cc}
a & b \\
b^{\ast} & a
\end{array} \right)$ and $V_2=\left( \begin{array}{cc}
c & d \\
d^{\ast} & c
\end{array} \right)$. A physical quantum system implies $a^2\ge |b|^2+\frac{1}{4}$, $c^2\ge |d|^2+\frac{1}{4}$ from uncertainty principle \cite{Simon:00}.

By expressing the characteristic function in terms of the output complex amplitudes $\beta_1,\beta_2$ using the transformation Eqs. \eqref{eq:BS}, \eqref{eq:M}, \eqref{eq:chara_f} the output covariance matrix is given by a unitary transformation of $V_{\text{in}}$ as
\begin{eqnarray}
V_{\text{out}}=U^{\dagger}(\theta,\varphi)V_{\text{in}}U(\theta,\varphi)= \left( \begin{array}{cc}
A & C \\
C^{\dagger} & B
\end{array} \right),
\end{eqnarray}
where $U(\theta,\varphi)$ is related to $M$ and it is given by
\begin{eqnarray}
U(\theta,\varphi) = \left( \begin{array}{cccc}
\cos\theta & 0 &-\sin\theta e^{i\varphi} & 0\\
0 & \cos\theta & 0 &-\sin\theta e^{-i\varphi}\\
\sin\theta e^{-i\varphi} & 0 & \cos\theta &0\\
0 &\sin\theta e^{i\varphi} & 0 & \cos\theta
\end{array} \right).\nonumber\\
\end{eqnarray}
The expressions of matrices $A$, $B$, and $C$ are given in Appendix A.

\subsection{Single-mode nonclassicality}
For any quantum state $\rho$, it can be represented in the Glauber-Sudarshan representation as
\begin{eqnarray}
\rho=\int P(\alpha,\alpha^{\ast})\ket{\alpha}\bra{\alpha}d^2\alpha,
\end{eqnarray}
where $\ket{\alpha}$ is a coherent state and $P(\alpha,\alpha^{\ast})$ is the Glauber-Sudarshan $P$ function defined as
\begin{eqnarray}
P(\alpha,\alpha^{\ast})=\frac{1}{\pi^2}\int e^{\frac{1}{2}|\beta|^2-i\beta \alpha^{\ast}-i\beta^{\ast}\alpha} \chi(\alpha,\alpha^{\ast})d^2\beta.
\end{eqnarray}
If the $P$ function of a quantum state is positive-definite, then the state is defined as a classical state. Otherwise, it is nonclassical. There are many nonclassicality quantifications proposed for a single-mode state \cite{Hillery:87,Lee:91,Dodonov:00,Marian:02, Asboth:05}. We first consider the nonclassicality depth \cite{Lee:91}. For a non-positive-definite $P$ function, a convolution of the $P$ function 
\begin{eqnarray}
R(\tau,\eta,\eta^{\ast})=\frac{1}{\pi\tau}\int e^{-1/\tau|\alpha-\eta|^2}P(\alpha,\alpha^{\ast})d^2\alpha
\end{eqnarray}
may become a positive-definite function as a classical probability distribution. For a given $P$ function, the minimum value of $\tau$ such that $R$ function becomes positive-definite is defined as the nonclassical depth. It is a measure of nonclassicality ranging between $0$ and $1$. Particularly, the nonclassicality depth is a continuous measure for a Gaussian state in the range of $[0,\frac{1}{2}]$. For the covariance matrix $V_1$, the nonclassical depth is given by
\begin{eqnarray}
\tau=\text{max}\{0,1/2-\lambda_{1\text{min}}\}
\end{eqnarray}
where $\lambda_{1\text{min}}=a-|b|$ is the minimum eigenvalue of $V_1$. Therefore, for any single-mode nonclassical Gaussian state, $\tau>0$ or $a-|b|<\frac{1}{2}$.

For a single-mode Gaussian state, the degree of squeezing can also be used as a quantification of nonclassicality \cite{Wolf:03}. If a Gaussian state is squeezed, the minimum uncertainty of its phase-space quadratures, which equals to the minimum eigenvalue $\lambda_{\text{min}}$ of its covariance matrix, is smaller than $\frac{1}{2}$ \cite{SZ}. Therefore, we consider the quantity $N_{\text{noncl}}=-\log_2(2\lambda_{\text{min}})$ as the nonclassicality of a single-mode Gaussian state. For a coherent state, $N_{\text{noncl}}=0$. For a pure squeezed state with squeezing parameter $r$ \cite{SZ}, we find $N_{\text{noncl}}=2r$. For a thermal state with average thermal photon number $n$, $N_{\text{noncl}}=-\log_2(2n+1)<0$.

\section{Extracting the nonclassicality completely}

\subsection{Calculation of the remaining nonclassicality}

Operation of the beam-splitter transforms the density matrix as $\rho_{12}=\mathcal{M}_{\scriptscriptstyle{\rm BS}} \rho^{\text{(in)}}_1 \otimes \rho^{\text{(in)}}_2 \mathcal{M}_{\scriptscriptstyle{\rm BS}}^\dagger$, where $\mathcal{M}_{\scriptscriptstyle{\rm BS}}$ is the BS operator and $\rho^{\text{(in)}}_{1,2}$ are the density matrices for the input states of the BS. In order to study the nonclassicality of each output modes, we define $\rho_1=\text{Tr}_2(\rho_{12})$ and $\rho_2=\text{Tr}_1(\rho_{12})$ for each of the output modes from the output state $\rho_{12}$. We further define a separable output system as $\tilde{\rho}_{12}=\text{Tr}_2(\rho_{12})\otimes\text{Tr}_1(\rho_{12})$, see Fig. \ref{fig:scheme}, where two-mode entanglement is wiped out \cite{PS1}.  We show that the covariance matrix of $\tilde{\rho}_{12}$ after the tracing operation on the output state is $\tilde{V}_{\text{out}}=\left( \begin{array}{cc}
A & 0 \\
0 & B
\end{array} \right)$, as should be expected. Covariance matrices for single-mode states ($A$, $B$) are unaffected. The derivation is provided in Appendix A.

\begin{figure}
\centering
\includegraphics[width=0.5\textwidth]{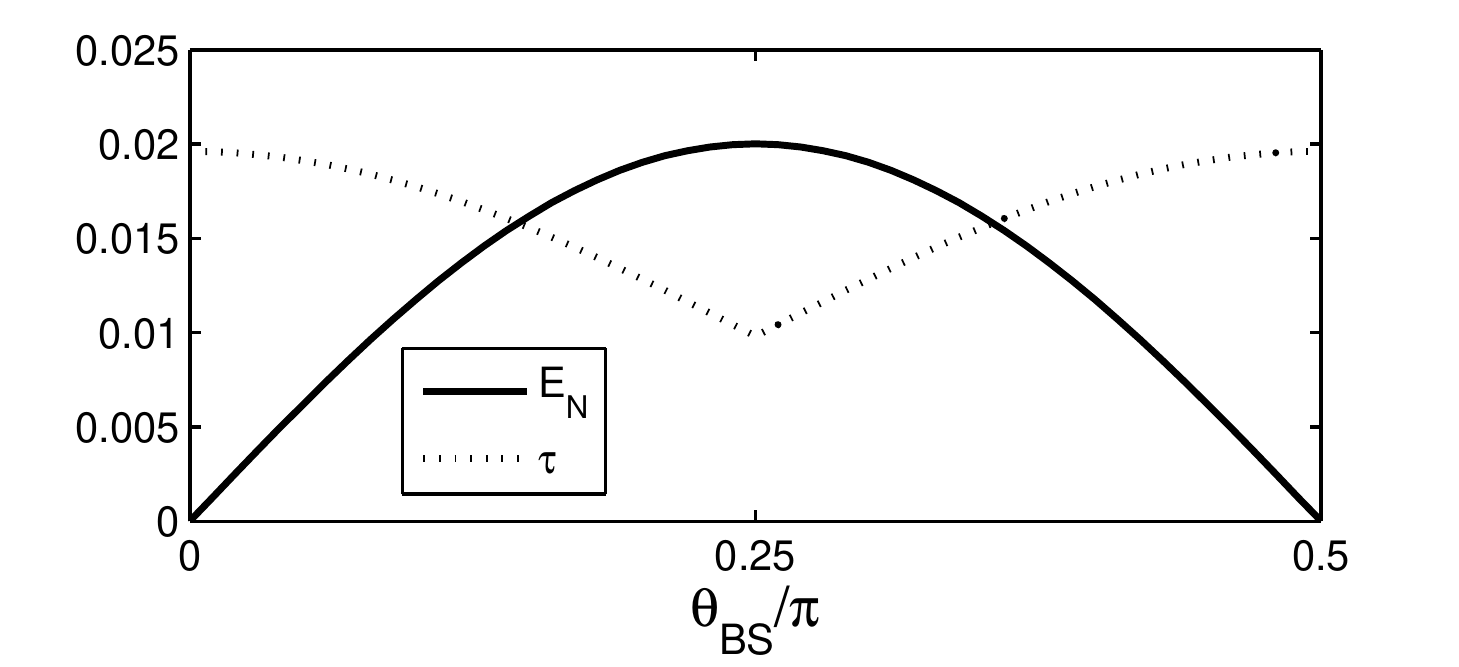}
\caption{Nonclassicality of the input single-mode state transforms in to two-mode entanglement in a beam-splitter. When the amount of extracted two-mode entanglement ($E_N$) increases, the nonclassicality in the output states ($\tau$) decreases. Not all of the nonclassicality could be converted to two-mode entanglement even for the optimum BS mixing angle.}
\label{fig:figENvstau}
\end{figure}

After this partial trace operation, we calculate the remaining nonclassical depth in the separable two-mode system, namely $\rho_1=\text{Tr}_2(\rho_{12})$ and $\rho_2=\text{Tr}_1(\rho_{12})$. We use the definition introduced in Ref.s~\cite{Li:06,serafini2005quantifying} to calculate the nonclassical depth for a two-mode Gaussian system. In Fig.~\ref{fig:figENvstau}, we plot the two-mode entanglement generated at the output of the BS and the remaining nonclassicality for different BS mixing angles. The nonclassicality is converted to two-mode entanglement in the BS, and there remains weaker nonclassicality as the strength of the two-mode entanglement increases.

\subsection{Depleting all nonclassicality}

\begin{figure}
\centering
\includegraphics[width=0.5\textwidth]{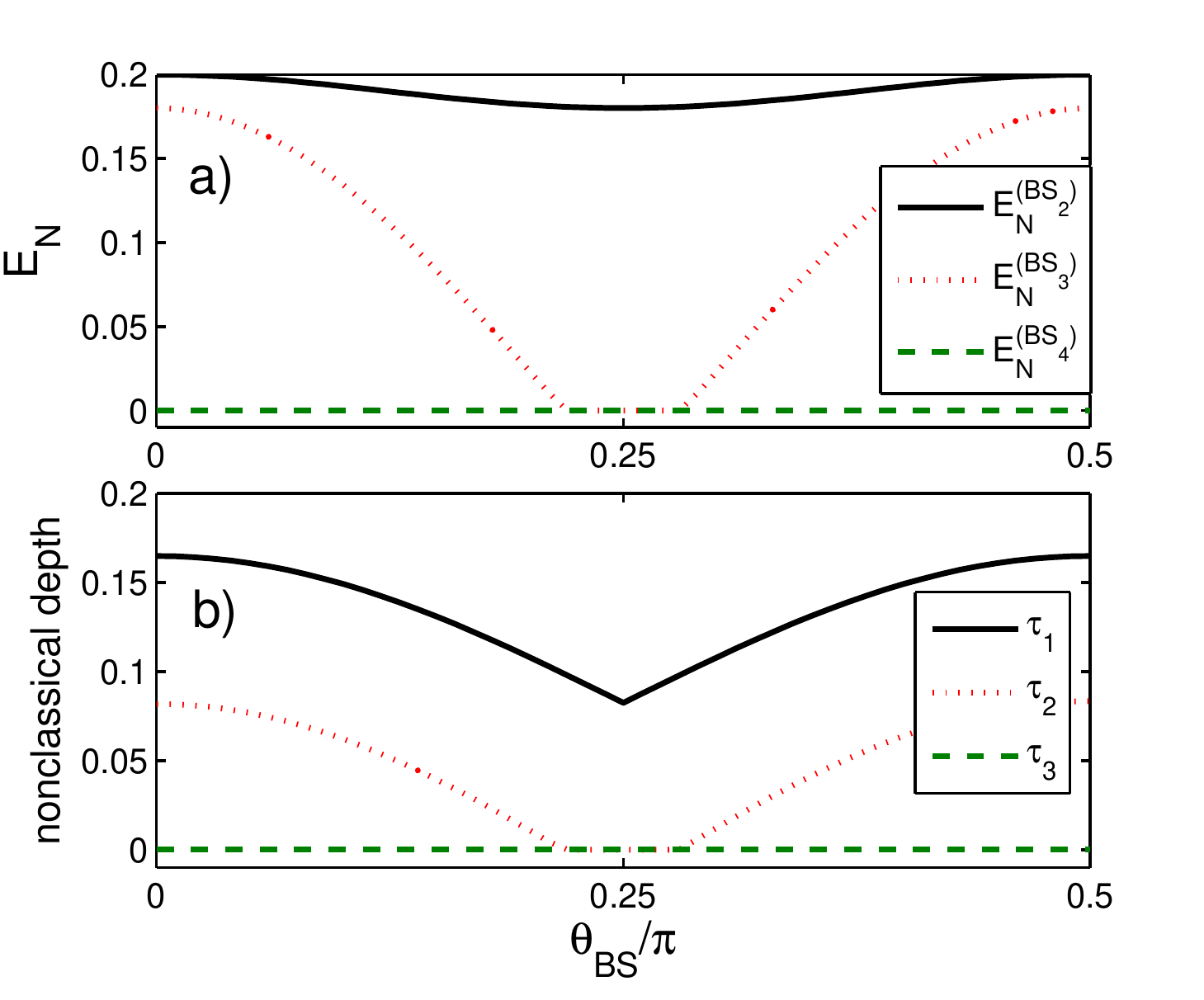}
\caption{(color online) The two mode output state of the first BS is partial traced and put in to a second BS, $\tilde{\rho}_{12}^{\scriptscriptstyle{\rm (BS_2)}}=\text{Tr}_2(\rho_{12}^{\scriptscriptstyle{\rm (BS_2)}})\otimes\text{Tr}_1(\rho_{12}^{\scriptscriptstyle{\rm (BS_2)}})$, in order to extract the remaining nonclassicality as entanglement. We repeat the same procedure for four BSs. The nonclassical depth before a BS, $\tau_{i-1}$, displays parallel behavior with the two-mode entanglement extracted from this BS, $E_N^{\scriptscriptstyle{{\rm (BS}_i {\rm )}}}$. Nonclassicality depletes at the fourth BS. }
\label{fig:figDepleting}
\end{figure}

We observe that not all of the nonclassicality is transformed to two-mode entanglement in Fig.~\ref{fig:figENvstau}. There remains nonclassicality in the two output modes. One naturally raises the question if we can transform all of the nonclassicality in to two-mode entanglement. For this reason, we put the separable state $\tilde{\rho}_{12}=\text{Tr}_2(\rho_{12})\otimes\text{Tr}_1(\rho_{12})$ ---after recording and wiping out the generated entanglement--- into following BSs \cite{PS1} to extract more of the entanglement potential. Since the two states are separable, i.e.  $\rho_1=\text{Tr}_2(\rho_{12})$ and $\rho_2=\text{Tr}_1(\rho_{12})$, we mix them in the second BS as $\rho_{12}^{{\scriptscriptstyle{\rm (BS_2)}}}=\mathcal{M}_{\scriptscriptstyle{\rm BS_2}} \tilde{\rho}_{12} \mathcal{M}_{\scriptscriptstyle{\rm BS_2}}^\dagger$. In Fig.~\ref{fig:figDepleting}a, we plot the extracted two-mode entanglement after the second BS, $E_N^{(\scriptscriptstyle{\rm BS_2})}$. We perform the partial trace operation again, $\tilde{\rho}_{12}^{\scriptscriptstyle{\rm (BS_2)}}=\text{Tr}_2(\rho_{12}^{\scriptscriptstyle{\rm (BS_2)}})\otimes\text{Tr}_1(\rho_{12}^{\scriptscriptstyle{\rm (BS_2)}})$, to obtain the remaining nonclassicality after the second BS, $\tau_2$, as plotted in Fig.~\ref{fig:figDepleting}b. We perform the similar procedure for two more BSs. There remains no nonclassicality after the third BS, $\tau_3=0$. This is also confirmed by placing a fourth BS where no two-mode entanglement can be extracted, $E_N^{\scriptscriptstyle{\rm (BS_4)}}=0$. Comparing Fig.~\ref{fig:figDepleting}a and Fig.~\ref{fig:figDepleting}b, we observe that the nonclassical depth before a BS, $\tau_{i-1}$, has parallel behavior with the two-mode entanglement extracted from this BS, $E_N^{\scriptscriptstyle{{\rm (BS}_i {\rm )}}}$.

Even though we calculated the nonclassical depth of a two mode state using the definition of Ref.s~\cite{Li:06,serafini2005quantifying} for a qualitative (behavior) comparison, in the preceding sections we arrive a more useful definition for quantitative purposes.

\section{Conservation relation of single-mode nonclassicality and two-mode entanglement}
We consider, in general, two single-mode Gaussian states mixed at a BS. The nonclassicalities of the input modes are $N_{\text{noncl}}^{\text{in}1}=-\log_2(2\lambda_{1\text{min}})$ and $N_{\text{noncl}}^{\text{in}2}=-\log_2(2\lambda_{2\text{min}})$, where $\lambda_{\text{min}i}$ is the minimum eigenvalue of $V_i$. After the BS, the nonclassicalities of the output modes are $N_{\text{noncl}}^{\text{out}1}=-\log_2(2\tilde{\lambda}_{1\text{min}})$ and $N_{\text{noncl}}^{\text{out}2}=-\log_2(2\tilde{\lambda}_{2\text{min}})$, where $\tilde{\lambda}_{1\text{min}}$ ($\tilde{\lambda}_{2\text{min}}$) is the minimum eigenvalue of matrix $A$ ($B$).

After the BS, two-mode entanglement can be generated from input nonclassical single-mode states. Since a BS is a linear device, which does not create extra nonclassicality, therefore we quantify the difference between input nonclassicality and output nonclassicality as the degree of two-mode entanglement generated via a BS. This quantity is denoted as 
\begin{eqnarray}
S_{\mathcal{N}}&=&N_{\text{noncl}}^{\text{in}1}+N_{\text{noncl}}^{\text{in}2}-N_{\text{noncl}}^{\text{out}1}-N_{\text{noncl}}^{\text{out}2}\nonumber\\
&=&\log_2\frac{\tilde{\lambda}_{1\text{min}}\tilde{\lambda}_{2\text{min}}}{\lambda_{1\text{min}}\lambda_{2\text{min}}}.
\label{eq:en_def}
\end{eqnarray}
We show in the following with several examples that this quantification of entanglement is equivalent to the logarithmic negativity \cite{Vidal:02}. Then we generalize this relation to class of Gaussian states mixed at a BS.

\begin{figure}[t]
\centering
\begin{subfigure}[t]{0.425\textwidth}
\includegraphics[width=1.0\textwidth]{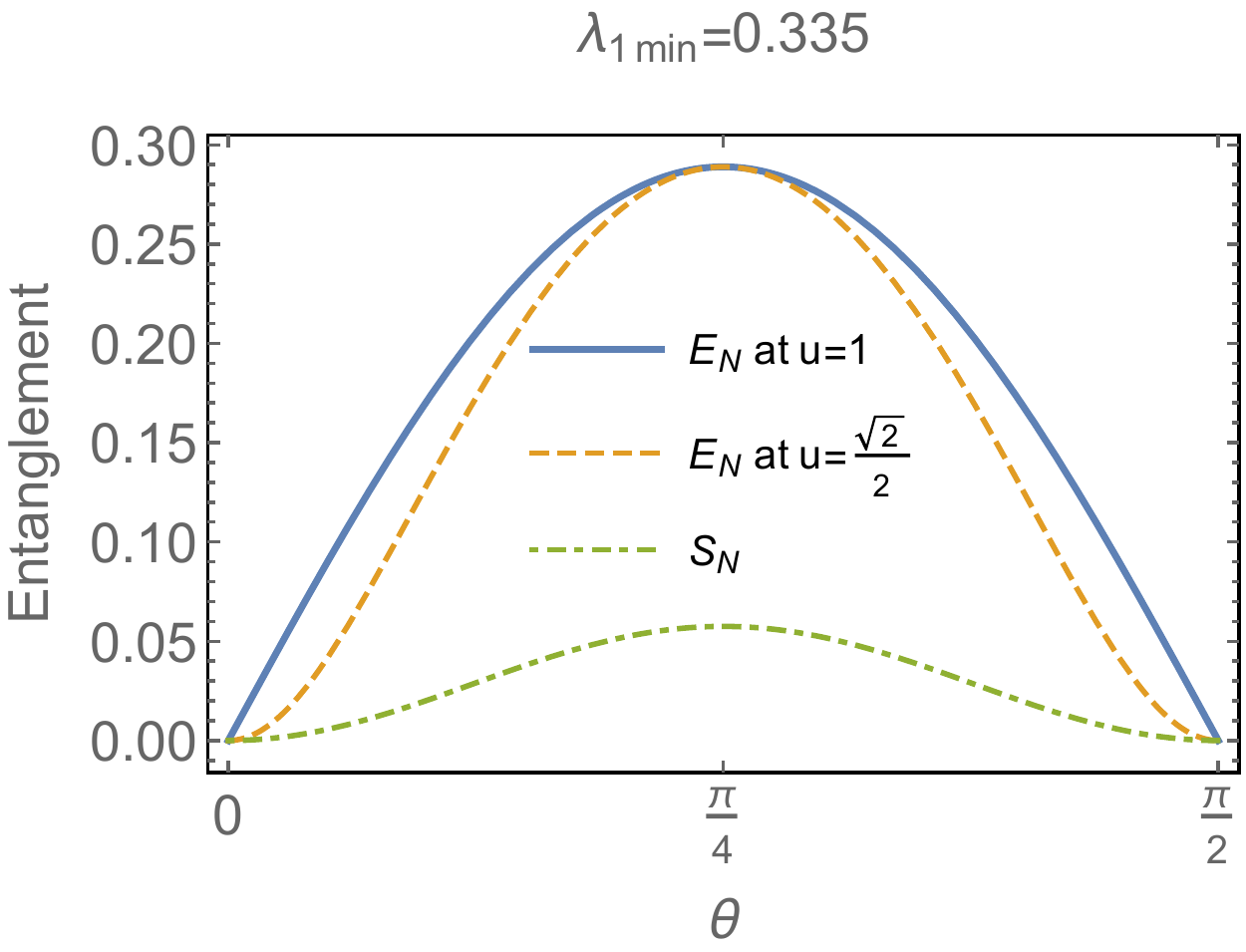}
\end{subfigure}
\begin{subfigure}[t]{0.425\textwidth}
\centering
\includegraphics[width=1.0\textwidth]{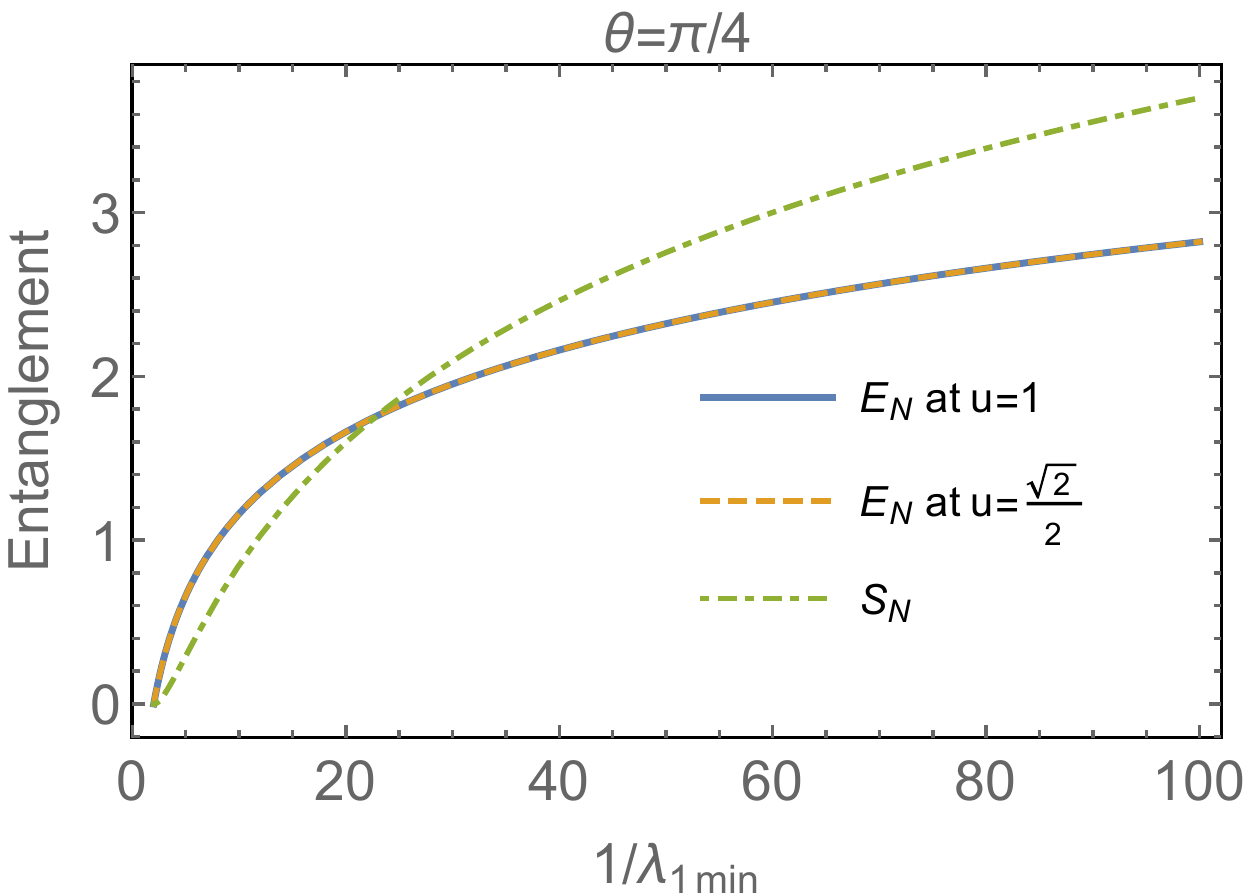}
\end{subfigure}
\caption{\label{fig:vacuum} (Color online). Output entanglement from mixing a nonclassical state with a vacuum at a BS quantified by two measures of degree of entanglement. (a) For constant nonclassicality, the relation of the degree of entanglement vs $\theta$ for different purities of the input nonclassical state. (b) At optimal BS angle, the monotonic relation of the degree of entanglement vs the inverse of the minimum eigenvalue.}
\end{figure}

\subsection{A single-mode nonclassical state mixing with a vacuum}
\subsubsection{Conservation relation of nonclassicality depth in a BS}
We first consider a simple case when a single-mode nonclassical Gaussian state is mixed with a vacuum at a BS. The covariance matrix of the two single-mode input state is given by $V_{\text{in}} = \left( \begin{array}{cc}
V_1 & 0 \\
0 & V_2
\end{array} \right),$ where $V_{1} = \left( \begin{array}{cc}
a & b \\
b^{\ast} & a
\end{array} \right)$ and $V_{2} = \left( \begin{array}{cc}
1/2 & 0 \\
0 & 1/2
\end{array} \right)$. Here $a$ is real and $b$ is complex in general. The eigenvalues of the matrix $V_1$ are $\lambda_{1\text{min}}=a- |b|$, $\lambda_{1\text{max}}=a+|b|$, and $u\equiv\frac{1}{2\sqrt{\lambda_{1\text{min}}\lambda_{1\text{max}}}}$ is the purity of the state \cite{Paris:03}. The eigenvalues of $V_2$ are $\lambda_{2\text{min}}=\lambda_{2\text{max}}=\frac{1}{2}$.

We show an interesting equation of nonclassical depth before and after the BS. Before the BS, the single-mode nonclassical state has a nonclassical depth $\tau$.  After mixing the nonclassical state with a vacuum at the BS, the nonclassical depth is calculated from matrices $A$ and $B$ for each output mode. The corresponding nonclassicality depths are given by $\tilde{\tau}_1=\tau\cos^2\theta$ and $\tilde{\tau}_2=\tau\sin^2\theta$. Therefore, we obtain
\begin{eqnarray}
\tau=\tilde{\tau}_1+\tilde{\tau}_2,
\end{eqnarray}
which is equivalent to 
\begin{eqnarray}
\lambda_{1\text{min}}+\lambda_{2\text{min}}=\tilde{\lambda}_{1\text{min}}+\tilde{\lambda}_{2\text{min}},
\label{eq:conservation1}
\end{eqnarray}
where $\tilde{\lambda}_{1\text{min}}$ and $\tilde{\lambda}_{2\text{min}}$ are minimum eigenvalues of $A$ and $B$, respectively. The conservation relation of nonclassicality depth holds for any nonclassical Gaussian state mixed with a vacuum state. Another interesting question arises: if the nonclassicality is conserved before and after the BS in such a way, where does the entanglement comes from after the BS?

\subsubsection{Conservation relation of nonclassicality and entanglement}
In the following, we show that the quantification of two-mode entanglement in Eq. \eqref{eq:en_def} is equivalent to the logarithmic negativity and therefore the conservation relation of single-mode nonclassicality and two-mode entanglement can be obtained.

A number of separability conditions \cite{Simon:00,Duan:00,Vidal:02,Hillery:06,Sun:09} have been proposed to test the entanglement of a bipartite system based on partial transposition \cite{ Peres:96,Horodecki:96}. For a two-mode Gaussian state, there are necessary and sufficient conditions \cite{Simon:00,Duan:00, Vidal:02} which can be used as measures for two-mode entanglement. Here we consider the logarithmic negativity defined in Ref. \cite{Vidal:02}. For the output covariance matrix $V_{\text{out}}$, the logarithmic negativity is given by \cite{Vidal:02}
\begin{eqnarray}
E_\mathcal{N}=\text{max}\left\{0,-\frac{1}{2}\log_2\left(S-\sqrt{S^2-16\text{Det}[V_{\text{out}}]}\right)\right\},
\label{eq:log_def}
\end{eqnarray}
where $S=2(\text{Det}[A]+\text{Det}[B]-2\text{Det}[C])$ and $\text{Det}[V_{\text{out}}]=\text{Det}[V_{\text{in}}]=\frac{1}{4}\lambda_{1\text{min}}\lambda_{1\text{max}}=\frac{1}{16u^2}$. Then we obtain the expression $S$ in terms of the nonclassical depth $\tau$, the purity $u$, and the BS angle $\theta$ as
\begin{eqnarray}
S&=&(1-\tau)\left(\frac{1}{2u^2(1-2\tau)}+\frac{1}{2}\right)\nonumber\\
&-&\tau \left( \frac{1}{2u^2(1-2\tau)}-\frac{1}{2} \right)\cos(4\theta)
\end{eqnarray}
From Eq. \eqref{eq:log_def}, we find that the condition for the two-mode Gaussian state to be entangled is 
\begin{eqnarray}
S>\frac{1}{2}+8\text{Det}[V_{\text{out}}].
\label{eq:entangle_con}
\end{eqnarray}
By rearranging the expression of $S$, we obtain an equivalent condition of Eq. \eqref{eq:entangle_con} as 
\begin{eqnarray}
\label{eq:entangle_con2}
&&S-\left(\frac{1}{2}+8\text{Det}[V_{\text{out}}]\right)\nonumber\\
&=&\mathcal{C}\left(\frac{(1-2\tilde{\tau}_1)(1-2\tilde{\tau}_2)}{1-2\tau}-1\right)>0,
\end{eqnarray}
where $\mathcal{C}=\frac{1}{u^2\tau}-\frac{1-2\tau}{\tau}\ge2$. Therefore, we find that the quantification of two-mode entanglement in Eq. \eqref{eq:en_def} is
\begin{eqnarray}
S_{\mathcal{N}}&=&\log_2\frac{(1-2\tilde{\tau}_1)(1-2\tilde{\tau}_2)}{1-2\tau}\nonumber\\
&=&\log_2\left[\frac{S-\left(\frac{1}{2}+8\text{Det}[V_{\text{out}}]\right)}{\mathcal{C}}+1\right].
\label{eq:entangle_measure}
\end{eqnarray}
As can be seen from the above expression, $S_{\mathcal{N}}>0$ is a necessary and sufficient condition for a two-mode Gaussian entanglement to exist, which is equivalent to the condition of the logarithmic negativity. When $S\le\frac{1}{2}+8\text{Det}[V_{\text{out}}]$, $S_{\mathcal{N}}\le0$ which gives us a quantitative measure of how far the system is away from entanglement. 

To see the validity of $S_{\mathcal{N}}$ numerically, we plot the degree of entanglement using both $S_{\mathcal{N}}$ and $E_{\mathcal{N}}$ in Fig. \ref{fig:vacuum} by either varying the transmittance or the single-mode nonclassicality. We observe similar qualitative trends of both the measures. We also observe that $S_{\mathcal{N}}$ is independent of the purity $u$ of the nonclassical state, while $E_{\mathcal{N}}$ increases with $u$ except for $\theta=0,\pi/4,\pi/2$.

Next we rewrite the relation Eq. \eqref{eq:en_def} to obtain a conservation relation between the initial nonclassicality before the BS and the sum of the degree of entanglement and the remaining nonclassicality after BS as
\begin{eqnarray}
N_{\text{noncl}}^{\text{in}1}+N_{\text{noncl}}^{\text{in}2}=N_{\text{noncl}}^{\text{out}1}+N_{\text{noncl}}^{\text{out}2}+S_{\mathcal{N}},
\label{eq:conservation2}
\end{eqnarray}
This is the main result of our paper. The initial single-mode nonclassicality equals to the sum of output single-mode nonclassicality and output two-mode entanglement generated via the BS. 

\begin{figure}[t]
\centering
\includegraphics[width=0.425\textwidth]{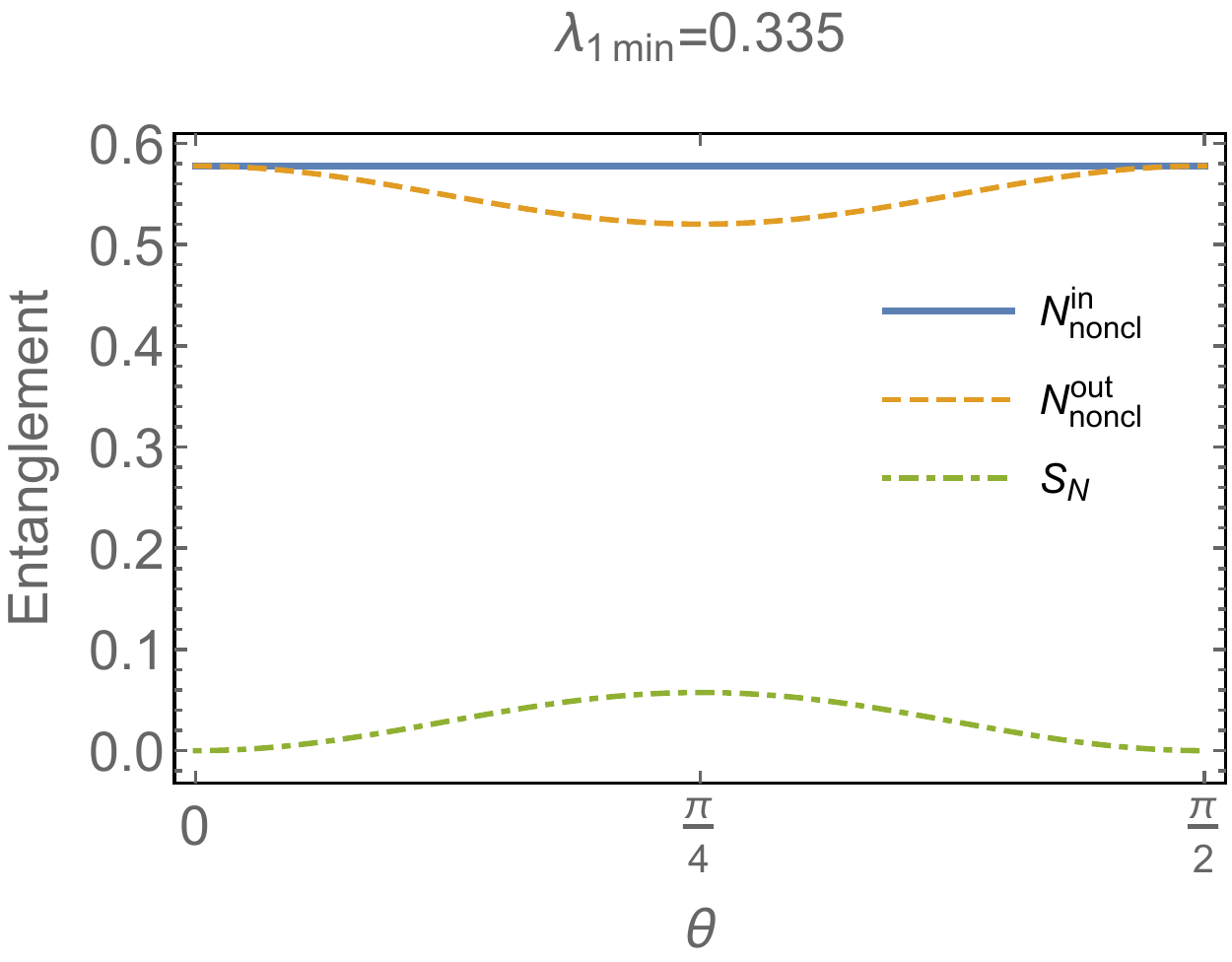}
\caption{\label{fig:vaccum_con} (Color online). $N_{\text{noncl}}^{\text{in}},N_{\text{noncl}}^{\text{out}}$ and $S_{\mathcal{N}}$ vs the BS angle $\theta$ for a nonclassical state mixing with a vacuum state.}
\end{figure}

We plot the curves of $N_{\text{noncl}}^{\text{in}}=N_{\text{noncl}}^{\text{in}1}+N_{\text{noncl}}^{\text{in}2}$, $N_{\text{noncl}}^{\text{out}}=N_{\text{noncl}}^{\text{out}1}+N_{\text{noncl}}^{\text{out}2}$ and $S_{\mathcal{N}}$ in Fig. \ref{fig:vaccum_con} to show this conservation relation using an example of a nonclassical state with $\lambda_{1\text{min}}=0.335$.

\subsection{A single-mode nonclassical Gaussian state mixing with a thermal state}
\begin{figure}[t]
\centering
\includegraphics[width=0.425\textwidth]{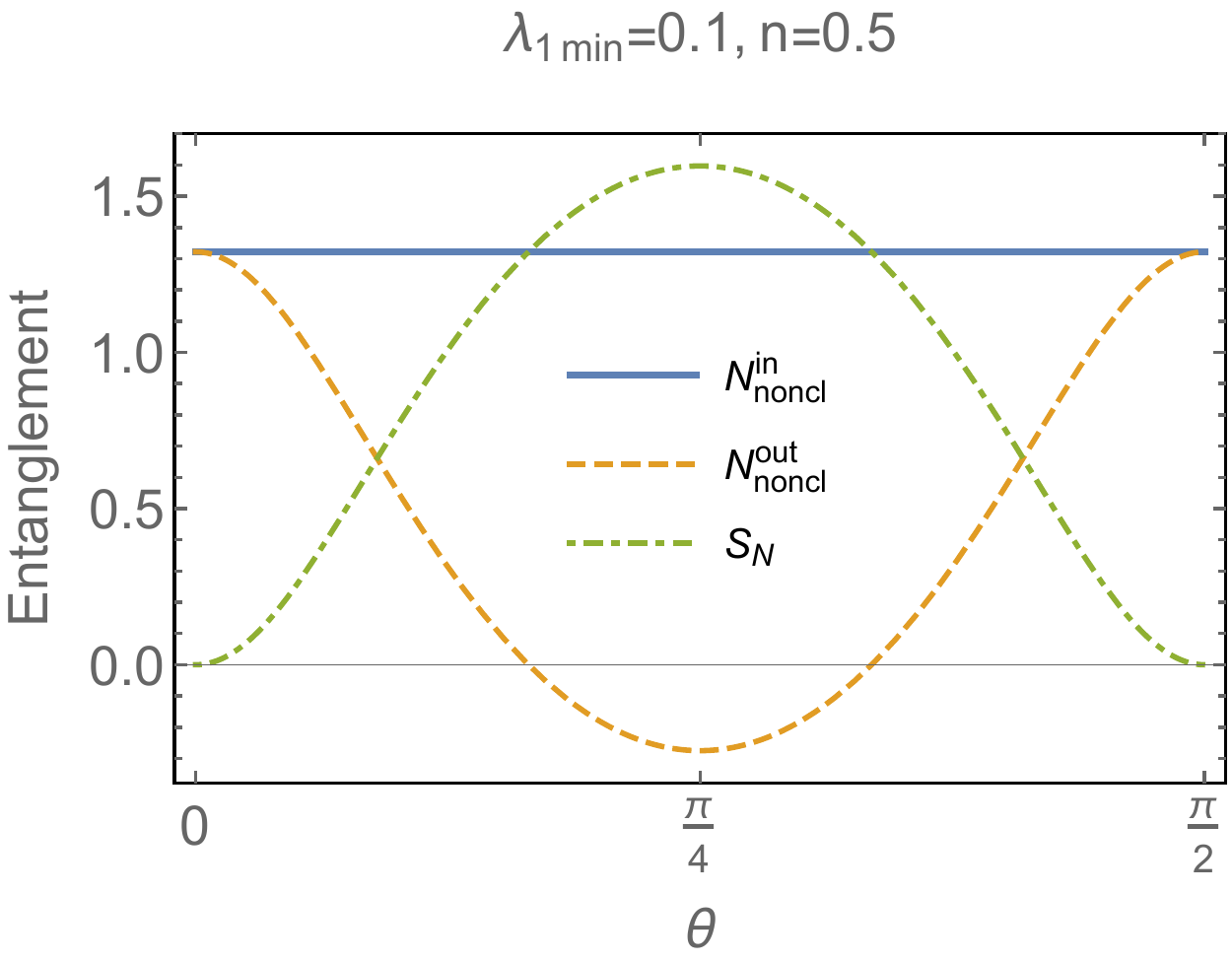}
\caption{\label{fig:thermal} (Color online). $N_{\text{noncl}}^{\text{in}},N_{\text{noncl}}^{\text{out}}$ and $S_{\mathcal{N}}$ vs the BS angle $\theta$ for a pure nonclassical state mixing with a thermal state.}
\end{figure}

Next we consider a pure nonclassical state with covariance matrix $V_1=\left( \begin{array}{cc}
a & b \\
b^{\ast} & a
\end{array} \right)$ mixing with a thermal state with a covariance matrix $V_{2} = \left( \begin{array}{cc}
n+\frac{1}{2} & 0 \\
0 & n+\frac{1}{2}
\end{array} \right)$, where $n$ is the average number of thermal photon. The input matrix is given by $V_{\text{in}} = \left( \begin{array}{cc}
V_1 & 0 \\
0 & V_2
\end{array} \right).$ Here the eigenvalues of $V_1$ satisfy $\lambda_{1\text{min}}\lambda_{1\text{max}}=a^2-|b|^2=\frac{1}{4}$ and the eigenvalues of $V_2$ are $\lambda_{2\text{min}}=\lambda_{2\text{max}}=n+\frac{1}{2}$. After the BS unitary transformation $U(\theta,\varphi)$, we obtain the expression of $S$ as 
\begin{eqnarray}
S&=&(\lambda_{1\text{min}}+n+\frac{1}{2})(\frac{1}{4\lambda_{1\text{min}}}+n+\frac{1}{2})\nonumber\\
&+&(\lambda_{1\text{min}}-n-\frac{1}{2})(\frac{1}{4\lambda_{1\text{min}}}-n-\frac{1}{2})\cos(4\theta).
\end{eqnarray}
The determinant of the output matrix $\text{Det}[V_{\text{out}}]=\text{Det}[V_{\text{in}}]=\frac{1}{4}(n+\frac{1}{2})^2$. We find the same expression as in the previous case
\begin{eqnarray}
S_{\mathcal{N}}&=&\log_2\frac{\tilde{\lambda}_{1\text{min}}\tilde{\lambda}_{2\text{min}}}{\lambda_{1\text{min}}\lambda_{2\text{min}}}\nonumber\\
&=&\log_2\left[\frac{S-\left(\frac{1}{2}+8\text{Det}[V_{\text{out}}]\right)}{\mathcal{C}}+1\right],
\end{eqnarray}
where in this case $\mathcal{C}=2(2n+1)\frac{1-2\lambda_{1\text{min}}(2n+1)}{2n+1-2\lambda_{1\text{min}}}$. In order to have $\mathcal{C}>0$ we require $2\lambda_{1\text{min}}(2n+1)<1$ which is the condition for entanglement to appear as discussed in Refs. \cite{Wolf:03,Tahira:09}. 

For the input nonclassical state, $N_{\text{noncl}}^{\text{in}1}=\log_2(\frac{1}{2\lambda_{1\text{min}}})>0$. For the input thermal state, $N_{\text{noncl}}^{\text{in}2}=-\log_2(1+2n)<0$, which means anti-squeezing and it is less nonclassical than a coherent state. Therefore the total nonclassicality of the input states are $N_{\text{noncl}}^{\text{in}}=N_{\text{noncl}}^{\text{in}1}+N_{\text{noncl}}^{\text{in}1}=-\log_2\left(2\lambda_{1\text{min}}(2n+1)\right)>0$. The remaining nonclassicality in the output states are $N_{\text{noncl}}^{\text{out}i}=\log_2(\frac{1}{2\tilde{\lambda}_{i\text{min}}})$ with $\tilde{\lambda}_{1\text{min}(2\text{min})}=\lambda_{1\text{min}(2\text{min})}\cos^2\theta+\lambda_{2\text{min} (1\text{min})}\sin^2\theta$.

With the definitions of  $S_{\mathcal{N}}$ and $N_{\text{noncl}}$, we obtain the same conservation relation as in Eq. \eqref{eq:conservation2} as
\begin{eqnarray}
N_{\text{noncl}}^{\text{in}1}+N_{\text{noncl}}^{\text{in}2}=N_{\text{noncl}}^{\text{out}1}+N_{\text{noncl}}^{\text{out}2}+S_{\mathcal{N}}.
\end{eqnarray}
We observe that, for optimal transfer of entanglement at the output, $\theta=\pi/4$ \cite{Wolf:03, Tahira:09}, $N_{\text{noncl}}^{\text{out}1},N_{\text{noncl}}^{\text{out}2}<0$ for $n\ge\frac{1}{2}$ since $\tilde{\lambda}_{1\text{min}(2\text{min})}=\frac{n}{2}+\frac{1}{4}+\frac{\lambda_{1\text{min}}}{2}>\frac{1}{2}$. Therefore, the output separable system $\tilde{\rho}_{12}$ has negative nonclassicality which means they are anti-squeezed.

We plot $N_{\text{noncl}}^{\text{in}},N_{\text{noncl}}^{\text{out}}$ and $S_{\mathcal{N}}$ in Fig. \ref{fig:thermal} and the relation $N_{\text{noncl}}^{\text{out}}+S_{\mathcal{N}}=N_{\text{noncl}}^{\text{in}}$ can be seen quantitatively from the figure. We also observe that more degree of entanglement than the input nonclassicality is obtained around $\theta=\pi/4$ due to the mixing with a thermal state.

\subsection{Generalization of the conservation relation} 
Next we show that the quantification of entanglement $S_{\mathcal{N}}$ can be generalized to any Gaussian states that satisfy the following two constraints:\\
(i) at least one of the input states is a pure state, i. e., the product of the two eigenvalues of the single-mode input state covariance matrix equals to $\frac{1}{4}$;\\
(ii) to make $\mathcal{C}$ positive, the eigenvalues of the input state satisfy the following conditions: $\lambda_{1\text{min}}<\lambda_{2\text{min}}<\lambda_{2\text{max}}<\lambda_{1\text{max}}$, or $\lambda_{2\text{min}}<\lambda_{1\text{min}}<\lambda_{1\text{max}}<\lambda_{2\text{max}}$, or $\lambda_{2\text{min}}=\lambda_{1\text{min}}$ and $\lambda_{1\text{max}}=\lambda_{2\text{max}}$.

We consider the case of mixing two general Gaussian states at a BS with at least one of them being a pure state. The input matrix is given by $V_{\text{in}} = \left( \begin{array}{cc}
V_1 & 0 \\
0 & V_2
\end{array} \right)$ where $V_1=\left( \begin{array}{cc}
a & b \\
b^{\ast} & a
\end{array} \right)$ and $V_2=\left( \begin{array}{cc}
c & d \\
d^{\ast} & c
\end{array} \right)$. Without loss of generality, we require that the product of the eigenvalues of $V_1$ satisfies $\lambda_{1\text{min}}\lambda_{1\text{max}}=a^2-|b|^2=\frac{1}{4}$. The eigenvalues of $V_2$ are $\lambda_{2\text{min}}=c-|d|$, $\lambda_{2\text{max}}=c+|d|$, and $\lambda_{2\text{min}}\lambda_{2\text{max}}\ge\frac{1}{4}$. When $|b|=0$ and $c-|d|<\frac{1}{2}$, this reduces to the first case of mixing a nonclassical state with a vacuum state. When $|d|=0$, this reduces to the second case of mixing a pure nonclassical state with a thermal state.
When $a-|b|,c-|d|<\frac{1}{2}$, this is the case of mixing a pure nonclassical state with another nonclassical state.

\begin{figure}[t]
\centering
\includegraphics[width=0.425\textwidth]{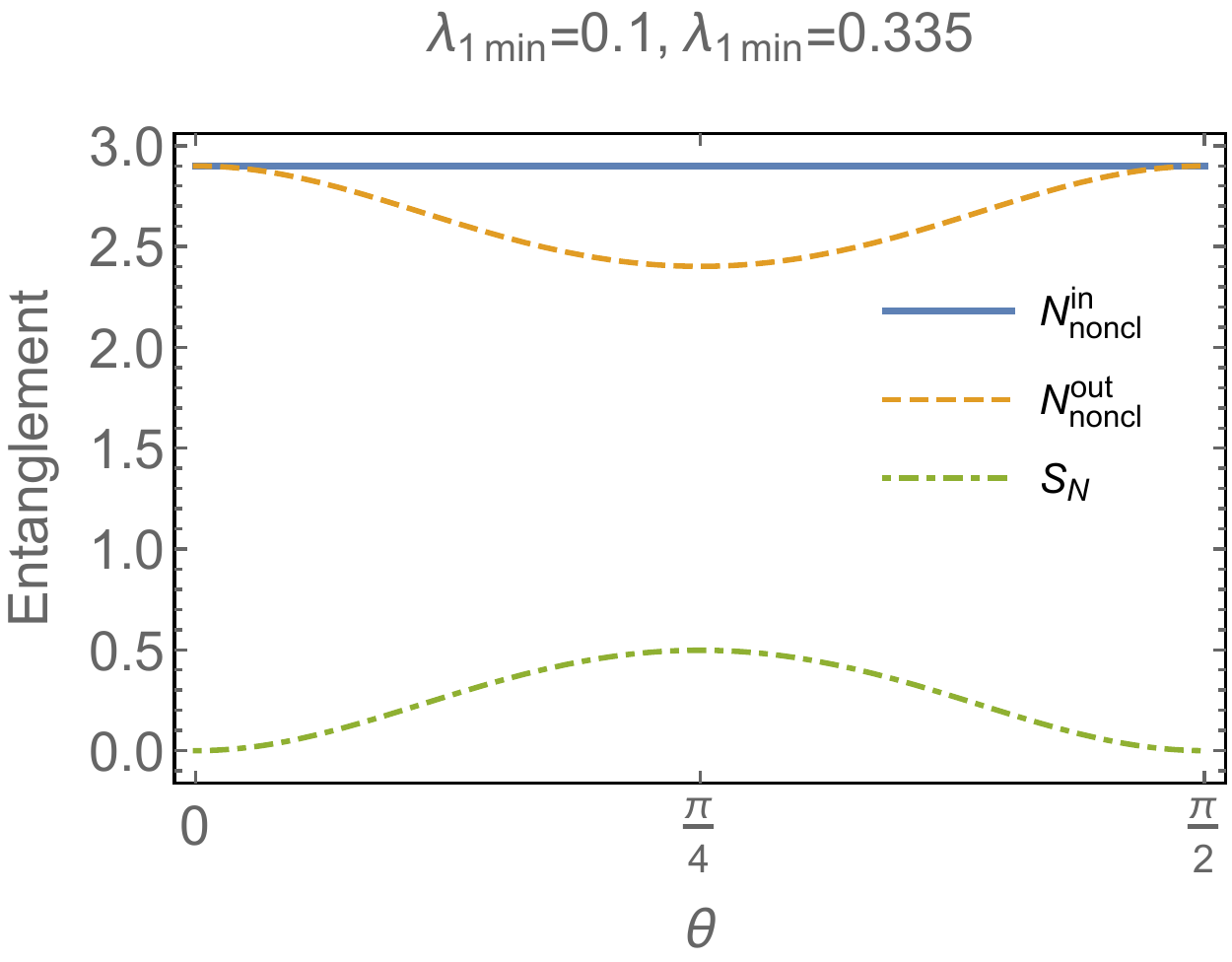}
\caption{\label{fig:nonclassical} (Color online). $N_{\text{noncl}}^{\text{in}},N_{\text{noncl}}^{\text{out}}$ and $S_{\mathcal{N}}$ vs the BS angle $\theta$ for a pure nonclassical state mixing with another nonclassical state.}
\end{figure}

As shown in Ref. \cite{Li:06}, the phase of the BS $\varphi$ plays a role since $b$ and $d$ are nonzero in general. For $b,d\ne0$, we require $\varphi=\text{arg}(b)/2-\text{arg}(d)/2$ to have a conservation relation of the minimum eigenvalues before and after the BS between the initial system and the output separable system, i. e.,
\begin{eqnarray}
\lambda_{1\text{min}}+\lambda_{2\text{min}}=\tilde{\lambda}_{1\text{min}}+\tilde{\lambda}_{2\text{min}}.
\end{eqnarray}
After the BS, the output matrix $V_{\text{out}}$ is given in the Appendix A. In general, we find
\begin{eqnarray}
S&=&(\lambda_{1\text{min}}+\lambda_{2\text{min}})(\lambda_{1\text{max}}+\lambda_{2\text{max}})\nonumber\\
&+&(\lambda_{1\text{min}}-\lambda_{2\text{min}})(\lambda_{1\text{max}}-\lambda_{2\text{max}})\cos(4\theta),
\end{eqnarray}
and
\begin{eqnarray}
\label{eq:entangle_con2}
S&-&\left(\frac{1}{2}+8\text{Det}[V_{\text{out}}]\right)\nonumber\\
&=&\mathcal{C}\left(\frac{\tilde{\lambda}_{1\text{min}}\tilde{\lambda}_{2\text{min}}}{\lambda_{1\text{min}}\lambda_{2\text{min}}}-1\right).
\end{eqnarray}
Here the expression of $\mathcal{C}$ is generalized as 
\begin{eqnarray}
\mathcal{C}&\equiv&8\lambda_{1\text{min}}\lambda_{2\text{min}}\frac{\lambda_{1\text{max}}-\lambda_{2\text{max}}}{\lambda_{2\text{min}}-\lambda_{1\text{min}}}.
\end{eqnarray}
Applying the constraint (ii), we find that $\mathcal{C}$ is positive-definite. Therefore the quantification $S_{\mathcal{N}}$ in Eq. \eqref{eq:en_def} is equivalent to the logarithmic negativity for any two single-mode Gaussian states satisfying the two constraints. Then we derive the conservation relation as
\begin{eqnarray}
N_{\text{noncl}}^{\text{in}1}+N_{\text{noncl}}^{\text{in}2}=N_{\text{noncl}}^{\text{out}1}+N_{\text{noncl}}^{\text{out}2}+S_{\mathcal{N}}.
\end{eqnarray}
The detailed derivation is provided in the Appendix A. We see that the sum of single-mode nonclassicality and two-mode entanglement is conserved before and after a BS under the unitary transformation $U(\theta,\varphi)$.

As an example of mixing two nonclassical states, a quantitative conservation relation between $N_{\text{noncl}}^{\text{in}},N_{\text{noncl}}^{\text{out}}$ and $S_{\mathcal{N}}$ is plotted in Fig. \ref{fig:nonclassical} for $\lambda_{1\text{min}}=0.1$ and $\lambda_{2\text{min}}=0.335$.

\begin{figure}[t]
\centering
\includegraphics[width=0.425\textwidth]{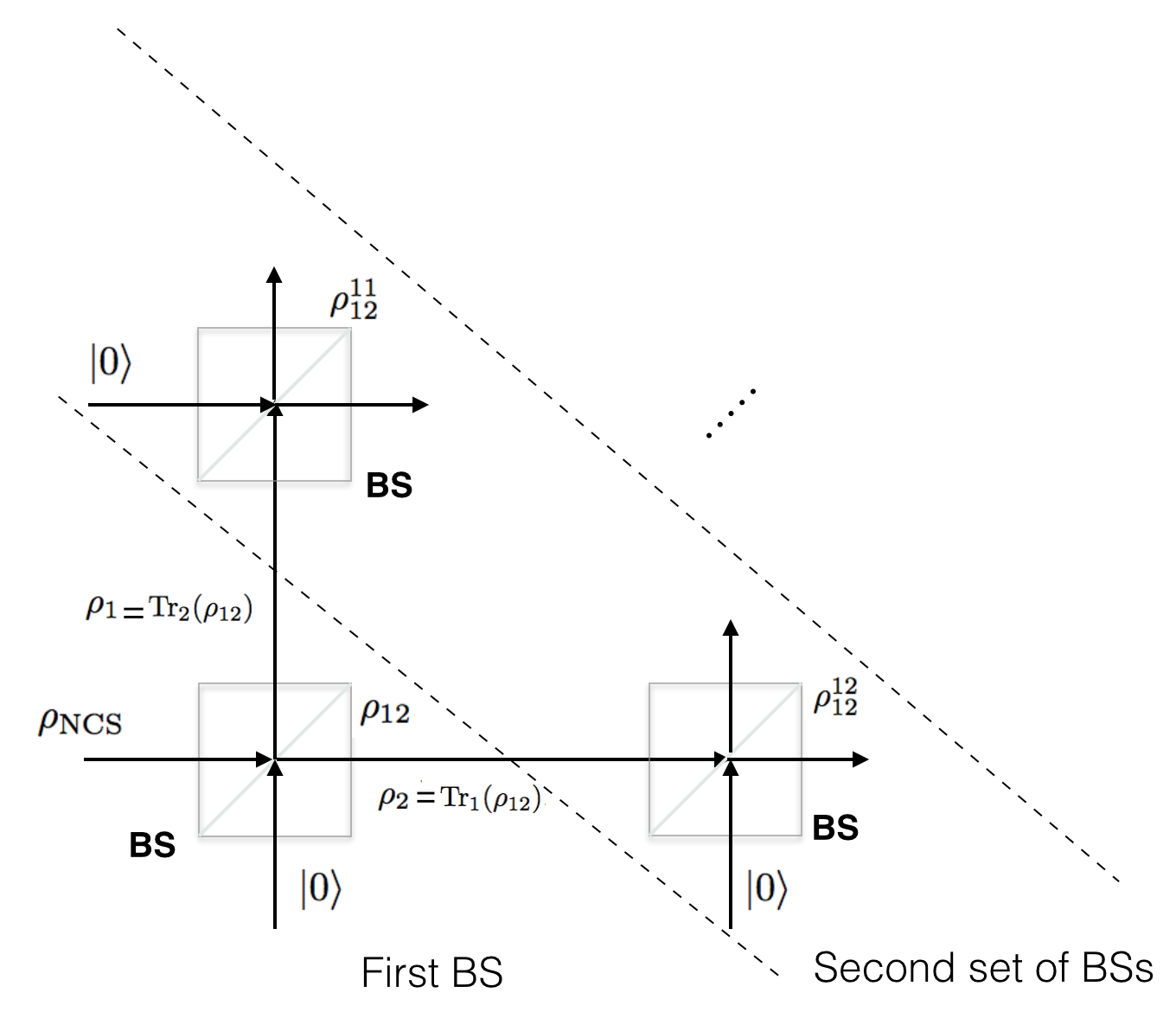}
\caption{\label{fig:scheme_ext} (Color online). A nonclassical state $\rho_{\text{NCS}}$ is mixed with a vacuum state at a BS, generating an output state $\rho_{12}$. Each output mode $\rho_1$ or $\rho_2$ is mixed with a vacuum state at another BS generating two sets of output bipartite Gaussian states $\rho_{12}^{11}$ and $\rho_{12}^{12}$. This process can be extended further with more BSs.}
\end{figure}

\subsection{Extension to infinite number of BSs}

As shown in Sec. IV A, after the first BS, we find that there is nonclassicality remaining in each of the output modes $\rho_1$ and $\rho_2$ for an initial nonclassical state mixed with a vacuum state. Therefore, we can send each output mode after the first BS, $\rho_1$ ($\rho_2$), to another BS mixing with a vacuum state to generate two sets of two-mode Gaussian entangled state as shown in Fig. \ref{fig:scheme_ext}. The generated entanglement and the remaining nonclassicality after the second set of BSs will be equal to the input nonclassicality before the set BSs. Then we can send each output mode after the second set of BSs to the third set of BSs mixing with a vacuum state separately. In this way, the nonclassicality is split further and after each set of BSs we create certain degree of entanglement. In each step, two conservation relations Eqs. \eqref{eq:conservation1}, \eqref{eq:conservation2} are satisfied. After infinite number of steps, there is some nonclassicality in each of the output single-mode state and by adding them up, we find total remaining nonclassicality is given by
\begin{eqnarray}
N_{\text{noncl}}^{\text{tot}}=(1-2\lambda_{1\text{min}})\log_2e,
\end{eqnarray}
which is independent on the angles of the BSs. Here $e$ is the Euler number. We can add all the entanglement generated in each step to obtain the total entanglement as
\begin{eqnarray}
S_{\mathcal{N}}^{\text{tot}}=\log_2\left(\frac{1}{2\lambda_{1\text{min}}}\right)-(1-2\lambda_{1\text{min}})\log_2e.
\end{eqnarray}
A simple proof is provided in the Appendix B. By extending our procedure to many sets of BSs, we find that both the quantifications of single-mode nonclassicality and two-mode entanglement are additive.

\section{Conclusion}
In this paper, we study the relation between the single-mode nonclassicality and two-mode entanglement created at a BS. We show that the input single-mode nonclassicality cannot be transferred completely into the output two-mode entanglement and there is remaining nonclassicality in the output modes. The more the generated entanglement, the less the remaining nonclassicality is. We use the logarithm of the minimum eigenvalue of a single mode covariance matrix (minimum uncertainty width) as its nonclassicality. 

We also define the difference between the input nonclassicality and the output nonclassicality as a degree for two-mode entanglement, which is generated from two single-mode Gaussian states mixed at a BS. This quantification has a qualitative correspondence with the logarithmic negativity. The sum of nonclassicality and entanglement is shown to be conserved before and after a BS using these quantifications. We generalize this conservation relation to a class of two-mode Gaussian states. Extension of many sets of BSs are discussed in the context of this conservation relation. Our work may stimulate a further interest in the unification of nonclassicality and entanglement.

\begin{acknowledgments}
This research is supported by an NPRP Grant (No. 7-201-1-032) from Qatar National Research Fund. M.E.T. acknowledge support from T\"{U}B\.{I}TAK-KAR\.{I}YER  Grant No.  112T927, T\"{U}B\.{I}TAK-1001  Grant No.  114F170 and  Hacettepe University BAP-6091 Grant No. 014G602002.
\end{acknowledgments}

\appendix
\section{Derivation of the conservation relation between entanglement and nonclassicality in a BS}
For a general input two-mode Gaussian covariance $V_{\text{in}} = \left( \begin{array}{cc}
V_1 & 0 \\
0 & V_2
\end{array} \right)$ with $V_1=\left( \begin{array}{cc}
a & b \\
b^{\ast} & a
\end{array} \right)$ and $V_2=\left( \begin{array}{cc}
c & d \\
d^{\ast} & c
\end{array} \right)$. After the BS, 
\begin{eqnarray}
V_{\text{out}}=U^{\dagger}(\theta,\varphi)V_{\text{in}}U(\theta,\varphi)= \left( \begin{array}{cc}
A & C \\
C^{\dagger} & B
\end{array} \right),
\end{eqnarray}
where the matrices $A$, $B$, and $C$ are given by
\begin{eqnarray}
A= \left( \begin{array}{cc}
a \cos^2\theta+c\sin^2\theta & b\cos^2\theta+d\sin^2\theta e^{2i\varphi} \\
 b^{\ast}\cos^2\theta+d^{\ast}\sin^2\theta e^{-2i\varphi} & a \cos^2\theta+c\sin^2\theta
\end{array} \right),\nonumber\\
\end{eqnarray}
\begin{eqnarray}
B= \left( \begin{array}{cc}
a \sin^2\theta+c\cos^2\theta & b\sin^2\theta e^{-2i\varphi}+d\cos^2\theta \\
 b^{\ast}\sin^2\theta e^{2i\varphi}+d^{\ast}\cos^2\theta & a \sin^2\theta+c\cos^2\theta
\end{array} \right),\nonumber\\
\end{eqnarray}
\begin{eqnarray}
C= \left( \begin{array}{cc}
(a-c)e^{i\varphi}& b e^{-i\varphi}-d e^{i\varphi} \\
 b^{\ast}e^{i\varphi}-d^{\ast}e^{-i\varphi} & (a -c)e^{-i\varphi}
\end{array} \right) \sin\theta\cos\theta.\nonumber\\
\end{eqnarray}
The elements of the covariance matrix are defined \cite{Simon:00} as $V_{\text{out}ij}=\frac{1}{2}\text{Tr}\big((v_iv_j+v_jv_i)\rho_{12}\big)$ where $v_i$ are position and momentum operators of the two-mode system defined as $v_1=x_1$, $v_2=p_1$, $v_3=x_2$ and $v_4=p_2$. 

For the output separable system $\tilde{\rho}_{12}=\text{Tr}_2(\rho_{12})\otimes\text{Tr}_1(\rho_{12})$, the elements of its covariance matrix are $\tilde{V}_{\text{out}ij}=\frac{1}{2}\text{Tr}\big((v_iv_j+v_jv_i)\tilde{\rho}_{12}\big)$. For $i=1,2$ and $j=3,4$, 
\begin{eqnarray}
\tilde{V}_{\text{out}ij}&=&\frac{1}{2}\big(\text{Tr}_1(v_i\rho_1)\text{Tr}_2(v_j\rho_2)+\text{Tr}_2(v_j\rho_2)\text{Tr}_1(v_i\rho_1)\big)\nonumber\\
&=&0
\end{eqnarray}
for zero-mean Gaussian states. For $i,j=1,2$
\begin{eqnarray}
\tilde{V}_{\text{out}ij}=\frac{1}{2}\text{Tr}_1\big((v_iv_j+v_jv_i)\rho_1\big)=V_{\text{out}ij}.
\end{eqnarray}
Similar relation holds for $i,j=3,4$. Therefore, we prove that $\tilde{V}_{\text{out}}= \left( \begin{array}{cc}
A & 0 \\
0 & B
\end{array} \right)$. The minimum eigenvalues of $A$ and $B$ are given by $\tilde{\lambda}_{1\text{min}(2\text{min})}=\lambda_{1\text{min}(2\text{min})}\sin^2{\theta}+\lambda_{2\text{min}(1\text{min})}\cos^2{\theta}$ using the phase condition $\varphi=\text{arg}(b)/2-\text{arg}(d)/2$. Therefore we obtain the first conservation relation between the minimum eigenvalues, i. e.,
\begin{eqnarray}
\lambda_{1\text{min}}+\lambda_{2\text{min}}=\tilde{\lambda}_{1\text{min}}+\tilde{\lambda}_{2\text{min}}.
\end{eqnarray}

We obtain the express of $S$ as
\begin{widetext}
\begin{eqnarray}
S&=&2(\text{Det}[A]+\text{Det}[B]-2\text{Det}[C])\nonumber\\
&=&2(a \cos^2\theta+c\sin^2\theta)^2-2(|b|\cos^2\theta+|d|\sin^2\theta)^2 +2(a \sin^2\theta+c\cos^2\theta)^2-2(|b|\sin^2\theta+|d|\cos^2\theta)^2\nonumber\\
&-&4(a-c)^2\sin^2\theta\cos^2\theta+4(|b|-|d|)^2\sin^2\theta\cos^2\theta\nonumber\\
&=&2(a^2-|b|^2+c^2-|d|^2)-8(a^2-|b|^2+c^2-|d|^2-2ac+2|b||d|)\sin^2\theta\cos^2\theta\nonumber\\
&=&2(a^2-|b|^2+c^2-|d|^2)-(a^2-|b|^2+c^2-|d|^2-2ac+2|b||d|)\big(1-\cos(4\theta)\big)\nonumber\\
&=&(\lambda_{1\text{min}}+\lambda_{2\text{min}})(\lambda_{1\text{max}}+\lambda_{2\text{max}})+(\lambda_{1\text{min}}-\lambda_{2\text{min}})(\lambda_{1\text{max}}-\lambda_{2\text{max}})\cos(4\theta)\nonumber\\
&=&(\lambda_{1\text{min}}+\lambda_{2\text{min}})(\frac{1}{4\lambda_{1\text{min}}}+\lambda_{2\text{max}})+(\lambda_{1\text{min}}-\lambda_{2\text{min}})(\frac{1}{4\lambda_{1\text{min}}}-\lambda_{2\text{max}})\cos(4\theta),
\end{eqnarray}
\end{widetext}
where we have used $\lambda_{1\text{min}}\lambda_{1\text{max}}=a^2-|b|^2=\frac{1}{4}$ and $\lambda_{2\text{min}(2\text{max})}=c\mp|d|$. With $\text{Det}[V_{\text{out}}]=\frac{1}{4}\lambda_{2\text{min}}\lambda_{2\text{max}}$, we obtain
\begin{eqnarray}
&&S-\left(\frac{1}{2}+8\text{Det}[V_{\text{out}}]\right)\nonumber\\
&&=\frac{\lambda_{2\text{min}}-\lambda_{1\text{min}}}{2\lambda_{1\text{min}}}(1-4\lambda_{1\text{min}} \lambda_{2\text{max}})\sin^2(2\theta).
\end{eqnarray}
Using $\mathcal{C}=2\lambda_{2\text{min}}\frac{1-4\lambda_{1\text{min}} \lambda_{2\text{max}}}{\lambda_{2\text{min}}-\lambda_{1\text{min}}}$ for $\lambda_{1\text{min}}\lambda_{1\text{max}}=\frac{1}{4}$, and $\frac{\tilde{\lambda}_{1\text{min}}\tilde{\lambda}_{2\text{min}}}{\lambda_{1\text{min}}\lambda_{2\text{min}}}-1=\frac{(\lambda_{2\text{min}}-\lambda_{1\text{min}})^2}{4\lambda_{1\text{min}}\lambda_{2\text{min}}}\sin^2(2\theta)$, we prove the equality 
\begin{eqnarray}
S&-&\left(\frac{1}{2}+8\text{Det}[V_{\text{out}}]\right)\nonumber\\
&=&\mathcal{C}\left(\frac{\tilde{\lambda}_{1\text{min}}\tilde{\lambda}_{2\text{min}}}{\lambda_{1\text{min}}\lambda_{2\text{min}}}-1\right).
\end{eqnarray} 

\section{Extension of infinite number of BSs}
When mixing a nonclassical state with a vacuum at a BS, we have
\begin{eqnarray}
\tau=\tilde{\tau}_1+\tilde{\tau}_2.
\end{eqnarray}
At the second set of BSs, we split the nonclassical depth further by mixing each subsystem with a vacuum state. Then we have
\begin{eqnarray}
\tilde{\tau}_1=\tilde{\tau}^{(11)}_1+\tilde{\tau}^{(11)}_2,
\end{eqnarray}
and 
\begin{eqnarray}
\tilde{\tau}_2=\tilde{\tau}^{(12)}_1+\tilde{\tau}^{(12)}_2.
\end{eqnarray}
After $m+1$ steps, there are $2^{m+1}$ single-mode Gaussian state generated and the nonclassical depth of each state is given by $\tilde{\tau}^{(mj)}_1,\tilde{\tau}^{(mj)}_2$, where $j=1,2,\dots,2^m$. As $m\rightarrow+\infty$, $\tilde{\tau}^{(mj)}_1,\tilde{\tau}^{(mj)}_2$ will be infinitesimal independent of the angles of the BSs at each step. Then the nonclassicality of each state is 
\begin{eqnarray}
N_{\text{noncl}}^{\text{out}(mj)1}=-\log_2(1-2\tilde{\tau}^{(mj)}_1)=2\tilde{\tau}^{(mj)}_1\log_2e.
\end{eqnarray}
The sum of all the nonclassicality after $m+1$ steps is 
\begin{eqnarray}
N_{\text{noncl}}^{\text{tot}}&=&\sum_{j=1}^{2^m}(N_{\text{noncl}}^{\text{out}(mj)1}+N_{\text{noncl}}^{\text{out}(mj)2})\nonumber\\
&=&2\sum_{j=1}^{2^m}(\tilde{\tau}^{(mj)}_1+\tilde{\tau}^{(mj)}_2)\log_2e\nonumber\\
&=&2\tau\log_2e=(1-2\lambda_{1\text{min}})\log_2e,
\end{eqnarray}
where the conservation relation of nonclassical depth is used. Using the conservation relation between nonclassicality and entanglement before and after a BS, we obtain
\begin{eqnarray}
S_{\mathcal{N}}^{\text{tot}}&=&N_{\text{noncl}}^{\text{in}}-N_{\text{noncl}}^{\text{tot}}\nonumber\\
&=&\log_2\left(\frac{1}{2\lambda_{1\text{min}}}\right)-(1-2\lambda_{1\text{min}})\log_2e.
\end{eqnarray}

\bibliography{bibliography}

\begin{thebibliography}{38}
\bibitem [{PS1()}]{PS1}%
  \BibitemOpen
  \href@noop {} {\bibinfo  {journal} {For developing the approach we present in
  this paper, such partial transposed states are not needed to be attained
  experimentally. Nevertheless, they can be experimentally obtained by mixing
  the opposite output ports from two identical BSs (both stand for the first
  BS) in a new (second) BS. One shall not perform any measurement on the
  not-mixed ports of the identical (first).}\ }\BibitemShut {NoStop}%
\end{thebibliography}

\end{document}